\documentclass[prl,aps,twocolumn,superscriptaddress,showpacs,nofootinbib]{revtex4-1}

\usepackage{amsmath}    
\usepackage{graphicx}   
\usepackage{color}
\usepackage[10pt]{moresize}

\usepackage{amsfonts}
\usepackage{amssymb}
\usepackage{amscd}
\usepackage{enumerate}
\usepackage{epsfig}
\usepackage{subfigure}
\usepackage{graphicx}
\usepackage{bm}
\usepackage{color}
\usepackage{epstopdf}

\begin{document}

\title{Bell Test Over Extremely High-Loss Channels:\\
Towards Distributing Entangled Photon Pairs Between Earth and Moon
}

\author{Yuan Cao}
\thanks{These authors contributed equally to this work}
\affiliation{Shanghai Branch, National Laboratory for Physical Sciences at Microscale and Department of Modern Physics, University of Science and Technology of China, Shanghai 201315, China}
\affiliation{Synergetic Innovation Center of Quantum Information and Quantum Physics, University of Science and Technology of China, Hefei, Anhui 230026, China}
\author{Yu-Huai Li}
\thanks{These authors contributed equally to this work}
\affiliation{Shanghai Branch, National Laboratory for Physical Sciences at Microscale and Department of Modern Physics, University of Science and Technology of China, Shanghai 201315, China}
\affiliation{Synergetic Innovation Center of Quantum Information and Quantum Physics, University of Science and Technology of China, Hefei, Anhui 230026, China}
\author{Wen-Jie Zou}
\affiliation{Shanghai Branch, National Laboratory for Physical Sciences at Microscale and Department of Modern Physics, University of Science and Technology of China, Shanghai 201315, China}
\affiliation{Synergetic Innovation Center of Quantum Information and Quantum Physics, University of Science and Technology of China, Hefei, Anhui 230026, China}
\author{Zheng-Ping Li}
\affiliation{Shanghai Branch, National Laboratory for Physical Sciences at Microscale and Department of Modern Physics, University of Science and Technology of China, Shanghai 201315, China}
\affiliation{Synergetic Innovation Center of Quantum Information and Quantum Physics, University of Science and Technology of China, Hefei, Anhui 230026, China}
\author{Qi Shen}
\affiliation{Shanghai Branch, National Laboratory for Physical Sciences at Microscale and Department of Modern Physics, University of Science and Technology of China, Shanghai 201315, China}
\affiliation{Synergetic Innovation Center of Quantum Information and Quantum Physics, University of Science and Technology of China, Hefei, Anhui 230026, China}
\author{Sheng-Kai Liao}
\affiliation{Shanghai Branch, National Laboratory for Physical Sciences at Microscale and Department of Modern Physics, University of Science and Technology of China, Shanghai 201315, China}
\affiliation{Synergetic Innovation Center of Quantum Information and Quantum Physics, University of Science and Technology of China, Hefei, Anhui 230026, China}
\author{Ji-Gang Ren}
\affiliation{Shanghai Branch, National Laboratory for Physical Sciences at Microscale and Department of Modern Physics, University of Science and Technology of China, Shanghai 201315, China}
\affiliation{Synergetic Innovation Center of Quantum Information and Quantum Physics, University of Science and Technology of China, Hefei, Anhui 230026, China}
\author{Juan Yin}
\affiliation{Shanghai Branch, National Laboratory for Physical Sciences at Microscale and Department of Modern Physics, University of Science and Technology of China, Shanghai 201315, China}
\affiliation{Synergetic Innovation Center of Quantum Information and Quantum Physics, University of Science and Technology of China, Hefei, Anhui 230026, China}
\author{Yu-Ao Chen}
\affiliation{Shanghai Branch, National Laboratory for Physical Sciences at Microscale and Department of Modern Physics, University of Science and Technology of China, Shanghai 201315, China}
\affiliation{Synergetic Innovation Center of Quantum Information and Quantum Physics, University of Science and Technology of China, Hefei, Anhui 230026, China}
\author{Cheng-Zhi Peng}
\affiliation{Shanghai Branch, National Laboratory for Physical Sciences at Microscale and Department of Modern Physics, University of Science and Technology of China, Shanghai 201315, China}
\affiliation{Synergetic Innovation Center of Quantum Information and Quantum Physics, University of Science and Technology of China, Hefei, Anhui 230026, China}
\author{Jian-Wei Pan}
\affiliation{Shanghai Branch, National Laboratory for Physical Sciences at Microscale and Department of Modern Physics, University of Science and Technology of China, Shanghai 201315, China}
\affiliation{Synergetic Innovation Center of Quantum Information and Quantum Physics, University of Science and Technology of China, Hefei, Anhui 230026, China}

\date{\today}

\begin{abstract}
Quantum entanglement was termed ``spooky action at a distance'' in the well-known paper by Einstein, Podolsky, and Rosen.
Entanglement is expected to be distributed over longer and longer distances in both practical applications and fundamental research into the principles of nature.
Here, we present a proposal for distributing entangled photon pairs between the Earth and Moon using a Lagrangian point at a distance of 1.28 light seconds.
One of the most fascinating features in this long-distance distribution of entanglement is that we can perform Bell test with human supply the random measurement settings and record the results while still maintaining space-like intervals.
To realize a proof-of-principle experiment, we develop an entangled photon source with 1 GHz generation rate, about 2 orders of magnitude higher than previous results.
Violation of the Bell's inequality was observed under a total simulated loss of 103 dB with measurement settings chosen by two experimenters.
This demonstrates the feasibility of such long-distance Bell test over extremely high-loss channels, paving the way for the ultimate test of the foundations of quantum mechanics.
\end{abstract}
\maketitle

Ever since it was first established, quantum mechanics has been the subject of intense debate about its intrinsic probabilistic and nonlocal nature, triggered by the Einstein-Podolsky-Rosen (EPR) paradox \cite{EPR_35}.
In particular, Bell’s inequality \cite{Bell_Ineq_64} manifests the contradiction between quantum mechanics and local hidden variable theories.
Increasing the entanglement distribution distance is of fundamental interest in studying its behavior.
Generally speaking, testing Bell's inequality while closing the known loopholes requires the entangled photon pairs to be distributed to distant parties. 
For example, if the distance could be extended far enough, it would be possible to use human free choice to address the ``freedom-of-choice'' loophole \cite{Bell:speakable:2004,Hall:measurementindependence:prl2010, RevModPhys.86.419,PhysRevA.93.032115,0264-9381-29-22-224011, PhysRevLett.106.100406}, or even allow the outcome to be observed by a conscious human observer to address the ``collapse locality loophole'' \cite{PhysRevA.72.012107,Manasseh2013, PhysRevLett.103.113601} in a Bell test.
This would also make it possible to test the effect of gravity on entanglement decorrelation \cite{0264-9381-29-22-224011}.
In terms of practical applications, extending entanglement to large scales would provide an essential physical resource for quantum information protocols such as quantum key distribution \cite{Bennett:BB84:1984,Ekert:QKD:1991,Bennett:BBM92:1992}, quantum teleportation \cite{BBCJPW_93, BPMEWZ_97}, and quantum networks \cite{Kimble2008}.
Distributing the entangled photon pairs as widely as possible is therefore an extremely important goal for a variety of reasons.

After more than 20 years of efforts, the maximum possible distance has been increased from a few meters to $\sim$ 100 km in optical fibers \cite{Honjo:08,Inagaki:13} or terrestrial free space \cite{yin:100kmtelenature:2012, Fedrizzi2009}, and has reached a limit on the earth due to photon loss in the channel. 
The most promising approach to dealing with this limitation is to use satellite- and space-based technologies. 
Excitingly, the first quantum science experiment satellite, ``Micius'', was successfully launched on 16 August, 2016 from Jiuquan, China.
Very recently, satellite-based quantum entanglement distribution over more than 1200 km has been demonstrated using ``Micius'' \cite{Yin1140}, taking the first step toward bringing Bell test to space. 

In this Letter, we design an experimental scheme for carrying out a Bell test between the earth and moon. 
To overcome the extremely high losses over the free-space optical links, we develop a new generation of ultra-high brightness entangled photon source and a high-time-resolution data acquisition system. 
It is worth noting that, at such a separation distance, the locality and freedom-of-choice loopholes can be completely closed, even using human free choice and human recorders instead of physical devices. 
We also implement a Bell test utilizing two human observers' choices with a high channel loss of 103 dB, demonstrating the scheme’s feasibility.

\begin{figure}
\centering
 \includegraphics[width=0.5\textwidth]{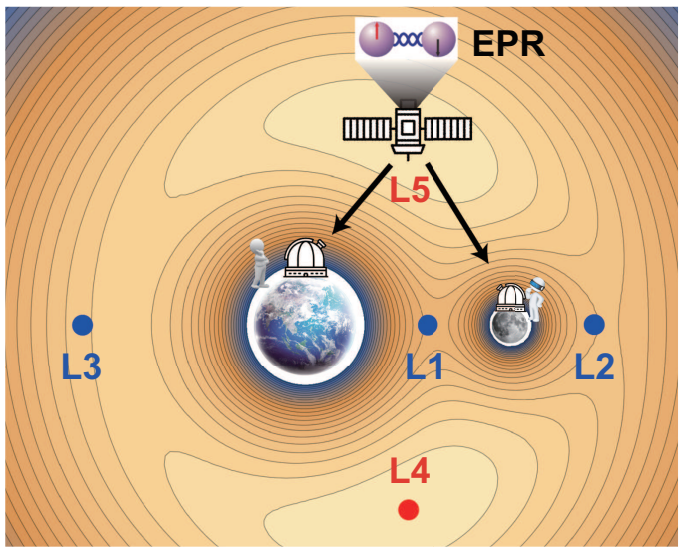}
\caption{(color online). 
Scheme for conducting a Bell test involving human free will. 
There are five Lagrangian points in the Earth-Moon system, denoted by L1, L2, L3, L4, and L5.
Since only L4 and L5 are stable, and have the most appropriate space arrangement of the five points, they were chosen for the position of the entanglement source satellite.
This satellite contains two telescopes, one aims at Moon and the other at Earth. 
Two large telescopes must also be built, one on or near Moon and one on Earth, to create the entanglement distribution channels.
}
\label{Fig:Lag}
\end{figure}

The Lagrangian points of Earth-Moon system are ideal places to put an entangled photon source, as shown in Fig.~\ref{Fig:Lag}.
We chose the L4 or L5 points for the entangled photon source because they are stable and have the most appropriate space arrangement in the five Lagrangian points.
The three points consisting of L4 (or L5), Earth, and Moon, form an approximately equilateral triangle with a side length of $3.8\times10^{5}$ km.
In this scheme, we define $A (B)$ and $a (b)$ as the events of Alice's (Bob's) measurement and setting choice, respectively.
The event where the entangled photon pairs are emitted by the entanglement source is defined as $S$.

\begin{figure}
{\includegraphics[width=0.5\textwidth]{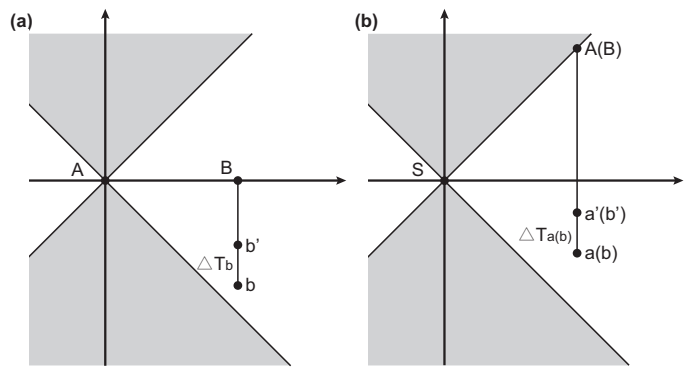}}
\caption{(color online). 
Space-time diagram for the scheme. 
The events $A (B)$, $a (b)$, and $a^\prime(b^\prime)$ represent Alice's (Bob's) measurement, setting choice and the measurement setting being prepared, respectively.
The event S represents the generation of the entangled photon pairs. 
(a). Closing the locality loophole. 
Due to the symmetry of $A (a)$ and $B (b)$, we only analyse $A$ and $b$ here without loss of generality.
The measurement events $A$ and $B$ happen nearly simultaneously.
To ensure that the interval between $A$ and $b$ is space-like, the delay between $b$ and $B$ should not exceed 1.28 s, the flight time required for light to traval from Moon to Earth.
Therefore, accounting for a delay of $\Delta T_b = 0.5 ~s$ due to human reaction time and system delay, photons that arrive within 0.78 s of the measurement setting being prepared are valid. 
(b). Closing the measurement independence loophole. 
$A(B)$ is on $E$'s light cones of $E$. 
For the interval between $E$ and $a (b)$ to be space-like, the delay between $a (b)$ and $A(B)$ must not exceed 2.56 s.
Thus, photons arriving within 2.06 s of the measurement setting being prepared are valid.
}
\label{Fig:spactime}
\end{figure}

This experimental scheme allows the locality loophole to be naturally closed, as shown in Fig.~\ref{Fig:spactime}(a).
Because the distances from the entanglement source to Moon and Earth are approximately equal, $A$ and $B$ can be considered to be simultaneous.
Clearly, $A$ and $B$ satisfy the space-like criterion.
In addition, we define $a^\prime (b^\prime)$ as the event where Alice's (Bob's) measurement setting be prepared and $\Delta T_a (\Delta T_b)$ as the delay between $a (b)$ and $a^\prime (b^\prime)$.
For the internal between $B$ and $a$ ($A$ and $b$) to be space-like, the time from $a (b)$ to $A (B)$ must be less than 1.28 s.
That is, once the measurement setting is ready, only photons that arrive within $1.28 ~s - \Delta T_a (\Delta T_b)$ are considered valid.
Second, Fig.~\ref{Fig:spactime}(b) shows the requirements for satisfying the freedom-of-choice assumption.
For the interval between $S$ and $a$ ($b$) to be space-like, the time from $a (b)$ to $A (B)$ must be less than 2.56 s, so the time from $a^\prime (B^\prime)$ to $A (B)$ needs to be less than $2.06 ~s - \Delta T_a (\Delta T_b)$.
In general, human reaction times are between 0.2 and 0.4 s \cite{BORGHI:1965vt}.
Including a system delay of 50 ms, defined as the time between the human observer pressing a key and $a^\prime (b^\prime)$, an upper bound of 0.5 s is reasonable for $\Delta T_a (\Delta T_b)$.
An exciting deduction from these two constraints is that it is feasible to perform a Bell test between Earth and Moon while avoiding the locality and freedom-of-choice loopholes, even when using humans to make the random selections and record the results, as long as we only consider photons that arrive within 0.78 s of the measurement setting being selected.

With current technologies, utilizing a 2.4-meter telescope (similar to the Hubble Space Telescope) in space and a 30-meter receiving telescope on Earth\cite{Sanders2013} (see Table I), the total loss in distributing the entanglement between Earth and Moon would be at least 100 dB, three orders of magnitude higher than in previous works \cite{yin:100kmtelenature:2012, cao:biasedbasis:2013, Yin2013Experimental}. 
This will be a big challenge, even for ground-based demonstrations. 
To overcome this extremely high entanglement distribution loss, two technologies will need to be developed.
\begin{table}\center
\begin{tabular}{|c|c|c|c|}
\hline & {Earth arm} & {Moon arm}\\
\hline ~~Geometry attenuation~~ & $~~~~~~32~dB~~~~~~$ & $~~~~~53.5~dB~~~~$ \\
\hline Atmosphere attenuate & $3~dB$ & $0~dB$ \\
\hline Optical components & $6~dB$ & $6~dB$ \\
\hline Detective efficiency & $0.5~dB$ & $0.5~dB$ \\
\hline Total loss & 41.5 dB & 60 dB\\
\hline \multicolumn{3}{|c|}{Two arms total loss: 101.5 dB}\\
\hline
\end{tabular}
\caption{
The estimated total loss for the entangled photon pairs between Earth and Moon can be divided into two components, namely Earth and Moon arms. 
With existing technology, a beam divergence of $3~\mu rad$ can be achieved for the satellite's transmitting telescope, while the diameter of the receiving telescopes can be up to 30 m and 2.4 m for Earth and Moon, respectively \cite{Sanders2013}.
For a transmission distance of $3.8\times10^{5}$ km, the geometry attenuations would be 32 dB and 53.5 dB respectively. 
The atmospheric attenuation, which only occurs for the Earth arm, would be approximately 3 dB. 
The optical components also contribute to the attenuation, mainly due to the fiber coupling efficiencies in the entanglement source and receiving telescopes (about 10 dB for the two arms) and the transmittance of the optical antennas (about 2 dB of two arms).
The efficiency of state-of-the-art single-photon detectors is at least 80\%\cite{0953-2048-25-6-063001, Marsili:2013fs}.
}
\label{tab:channalloss}
\end{table}

First, employing the same entanglement source used in previous experiments \cite{Scheidl:freedomofchoice:pnas2010,yin:100kmtelenature:2012,Yin:testspeed:prl2013} would mean it might take years to perform the Bell test \cite{SM}, which would pose a significant challenge. 
Instead, we have created a new type of quantum entanglement source, based on a Type-0 periodically-poled potassium titanyl phosphate (PPKTP) crystal and a Sagnac interferometer.
This can generate $1.0\times10^9$ entangled photon pairs per second, about two orders of magnitude higher than the sources in Ref.~\cite{Scheidl:freedomofchoice:pnas2010,yin:100kmtelenature:2012,Yin:testspeed:prl2013,Yin1140}.
Under certain conditions, the generation rate of a spontaneous parametric down-conversion (SPDC) process can be approached as \cite{Grice:1997ht,Grice:2001jc},
\begin{equation}
\label{ }
G\propto{[\chi_{eff}^{(2)}]}^2\cdot\frac{1}{|n_s^\prime-n_i^\prime|}.
\end{equation}
Here, $\chi_{eff}$ stands for the efficient second-order non-linear coefficient, while $n_s^\prime$ and $n_i^\prime$ are the group refractive indices of signal and idler photons in the PPKTP crystal, respectively.
The first and second terms represent the relative spectral intensity and spectral width, respectively.
In an SPDC process going from a 405 nm pump laser to approximately 810 nm parametric photons, the relative spectral intensity and spectral width are both higher for Type-0 than Type-II, leading to a generation rate that is $\sim$ 2 orders of magnitude higher \cite{SM}.

Second, a much higher-resolution time-to-digital converter (TDC) system must be developed if we are to achieve a sufficiently high signal-to-noise ratio at this higher entangled photon pair generation rate.
Based on the field-programmable-gate-arrays (FPGAs) carry chains used as tapped delay lines \cite{Shen:2013dc}, a homemade TDC, which has a full width at half maximum (FWHM) time resolution of approximately 60 ps for measuring coincident events in two channels \cite{Qi:2014gg}, is employed to record the photon detection results.
\begin{figure*}[!t]\center
\resizebox{14cm}{!}{\includegraphics{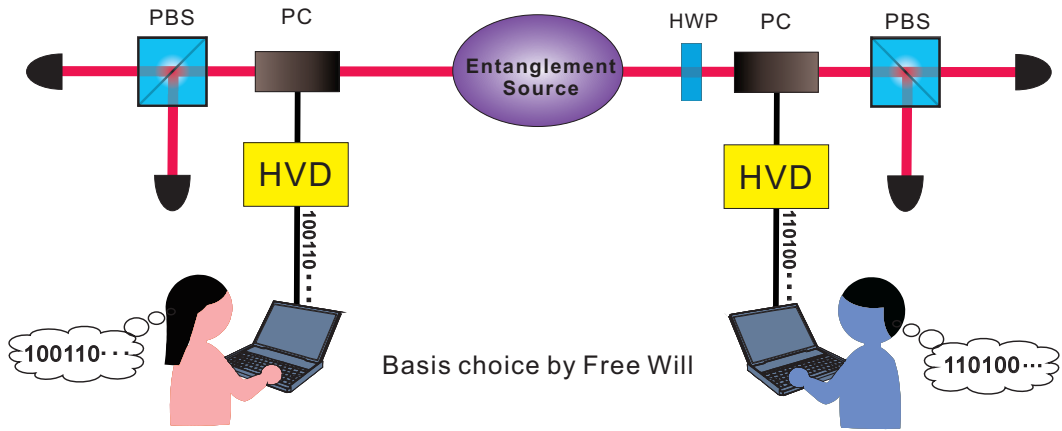}}
\caption{(color online). 
Simulated experiment setup. 
Entangled photon pairs are sent to two independent detection modules to perform the collective measurement. 
In each detection module, a Pockels Cell (PC) is used to select the basis chosen by the experimenter.
After being split using a polarized beam splitter (PBS), the photons are coupled to two single-mode fibers for detection.
To measure the CHSH inequality, the PCs are aligned at $22.5^\circ$ and an extra half-wave plate (HWP) at $11.25^\circ$ is placed in front of one of them. 
HVD: high-voltage driver.
}
\label{Fig:Setup}
\end{figure*}

Even though this scheme is feasible with current techniques, building a satellite equipped with an ultra-high-brightness entangled photon source and a large telescope on or near Moon are still enormous challenges.
Here, we perform a proof-of-principle experiment to demonstrate the feasibility via ultra-attenuating channel.
With two independent human observers choosing the bases, the CHSH inequality was measured.

As shown in Fig.~\ref{Fig:Setup}, signal and idler photons generated by a Type-0 PPKTP entanglement source were sent in different directions and measured by separate detection modules to distinguish their polarizations.
A large attenuation was applied to both arms to simulate the high losses of the satellite-Earth and satellite-Moon channels.
Each detection module was equipped with a Pockels Cell (PC) to select the basis, and a polarized beam splitter (PBS) and two single-mode fibers coupled single-photon detectors (SPDs) to measure the polarization.
The PC contains two potassium dihydrogen phosphate (KDP) crystals with half-wave voltage of $V_{\frac{\pi}{2}}\approx760 $ V @ $780$ nm and $V_{\frac{\pi}{2}}\approx820$ V @ $842$ nm.
When applied with the half-wave voltage, the PC became equivalent to a half-wave plate (HWP) and could thus rotate the photos's polarization.
By pressing on a keyboard, each human observer could manipulate a customized high-voltage driver to output either $V_{0}= 0$ V or the corresponding PC's half-wave voltage.
The delay between the key being pressed and the PC reaching the desired voltage is less than 50 ms. 
The photons transmitted through or reflected on the PBS are coupled to two single-mode fibers and detected by two Si SPDs with a jitter of approximately 40 ps and a quantum efficiency of 10\%.
Finally, the detector outputs are recorded by the custom TDC.
In total, the system's time resolution is approximately 82 ps (FWHM).

To conduct the simulated experiment, an observer sat near each of the two detection modules.
Each press a key board with a frequency of 2-4 Hz to select measurement bases without communicating to each other or negotiating a strategy.
The optical axes of the two PCs were rotated to $22.5^\circ$.
Each photon suffered a total loss of 51.5 dB, including 38.5 dB introduced by the attenuator, 3 dB by the single-mode fiber's coupling efficiency, and 10 dB by the SPDs' detection efficiency.
The total loss for a pair of entangled photons was therefore 103 dB.
It is reasonable to include the SPDs' detection efficiency in the simulated attenuation here, because a state-of-the-art SPD can achieve a quantum efficiency of 80\% with a jitter of 40 ps \cite{0953-2048-25-6-063001, Marsili:2013fs}.
The S-value for the CHSH inequality was measured to be $S=2.28\pm 0.061$ in 3 hours, with 537 coincidences.

The idea of using human free will to decide the measurement settings used for Bell test may originate from Bell himself \cite{Bell:speakable:2004,SM}.
If free will is assumed to exist, it naturally becomes a promising candidate for a stochastic source, due to its intrinsic attributes of freedom and independence. 
The ``freedom-of-choice'' loophole is receiving increasing attention in the field, and some researchers have suggested that human free will might be an effective means of addressing that loophole \cite{RevModPhys.86.419,PhysRevA.93.032115,Scheidl:freedomofchoice:pnas2010,0264-9381-29-22-224011,PhysRevLett.106.100406,Hall:measurementindependence:prl2010}, although this issue is still controversial mainly due to its imperfect randomness \cite{Wagenaar:1972hw}.
For example, nine scientific research institutions around the world came together to conduct a worldwide experiment called ``the big Bell test'' using human randomness on November 30, 2016 \cite{tbbt}.
However, given that the earth is only $\sim$43 light-ms in diameter, human choices are too slow to allow the separation of the measurements and entanglement source to be space-like.
The scheme proposed in this Letter offers a practical approach to solving this problem.

It should be noted that another way to address this loophole is to use cosmological signals coming from distant regions of space to determine the measurement settings. 
This would allow any local-realist model to be a limited space-time region, e.g., no more recently than approximately 600 years ago \cite{Handsteiner2017Cosmic}. 
Nonetheless, because this method requires detectable signals, even utilizing the Cosmic Microwave Background would only extend this back to 380,000 years after the Big Bang \cite{PhysRevLett.112.110405}. 
In addition, some additional assumptions must be introduced when using this method, such as the hidden variable model not influencing the energy, momentum \cite{Handsteiner2017Cosmic} or signal arrival times \cite{PhysRevLett.118.140402}.

It is also worth noting that the other man-made device that could be replaced by a human in a Bell test is the detector.
Specifically, it may be possible to detect entangled photons with the naked eyes \cite{PhysRevA.78.052110} and record results by human observers \cite{PhysRevA.72.012107}, involving human consciousness more deeply in the Bell test. 
In the current scheme, the entangled photons are detected directly after distribution, so the heralding efficiency is extremely low on both side.
This is a problem: since humans can only detect or record a few photons per second, it would take an unacceptable long time to obtain a sufficient number of coincidence events.
However, using an event-ready scheme \cite{PhysRevLett.71.4287,PhysRevLett.91.110405} and a quantum memory technique \cite{Yang:2016du} could increase the heralding efficiency significantly, making it possible to introduce human recorders.
At the meantime, such a proposal could even close the ``fair sampling loophole'' and realize a loophole-free Bell test. 
In addition, electroencephalogram (EEG) devices can predict human choices $\sim$0.1 s before the corresponding muscle actions \cite{Hardy2017}, which could be a useful way of slightly relaxing the space-like interval condition.

We also want to emphasize here that even though the natures of consciousness and free will are still unresolved problems, they can be treated with a more open and more scientific attitude with the continuous progress of technology in the fields of neuroscience and quantum physics \cite{Brembs930}.

In this work, we have proposed a scheme for conducting Bell test between Earth and Moon that would simultaneously address the measurement independence and locality loopholes, even when involving conscious human minds with free will.
We have estimated the total loss for this scheme and analyzed the space-time relationships.
To overcome the extremely high loss, we have realized a quantum entanglement source with a generation rate of approximately 1 GHz.
Using this ultra-high-brightness source, we were able to conduct an experiment where a quantum entanglement distribution with a loss of 103 dB was divided between two human observers, who selected the measurement bases.
These results demonstrate that it is feasible to handle such long-range entanglement distributions, potentially providing a new way to address these fascinating issues.

We acknowledge W.-Q. Cai for insightful discussions. This work has been supported by CAS Center for Excellence and Synergetic Innovation Center in Quantum Information and Quantum Physics, Shanghai Branch, University of Science and Technology of China, by the National Fundamental Research Program (under grant no. 2013CB336800), the National Natural Science Foundation of China, and by the Strategic Priority Research Program on Space Science, the Chinese Academy of Sciences.

\bibliographystyle{apsrev4-1}

%

\end{document}